# Beam dynamics studies at DAΦNE: from ideas to experimental results

**M Zobov on behalf of the DAΦNE Team**

LNF-INFN, Via Enrico Fermi 40, 00044 Frascati (Rome), Italy

E-mail: mikhail.zobov@lnf.infn.it

**Abstract**. DAΦNE is the electron-positron collider operating at the energy of Φ-resonance, 1 GeV in the center of mass. The presently achieved luminosity is by about two orders of magnitude higher than that obtained at other colliders ever operated at this energy. Careful beam dynamic studies such as the vacuum chamber design with low beam coupling impedance, suppression of different kinds of beam instabilities, investigation of beam-beam interaction, optimization of the beam nonlinear motion have been the key ingredients that have helped to reach this impressive result. Many novel ideas in accelerator physics have been proposed and/or tested experimentally at DAΦNE for the first time. In this paper we discuss the advanced accelerator physics studies performed at DAΦNE.

## 1. Introduction

DAΦNE is one of the seven charged particle colliders operating in the world. It is an electron-positron collider working at the energy of the Φ resonance (1.02 GeV c.m.) aimed at providing a high rate of K mesons production [1, 2]. The DAΦNE accelerator complex (see Fig 1) consists of two main rings and an injection system composed of a full energy linear accelerator, a damping/accumulator ring and transfer lines. The presently achieved peak luminosity of $4.5 \times 10^{32}$ cm$^{-2}$s$^{-1}$ is by about two orders of magnitude higher than that obtained at other colliders ever operated at the same energy.

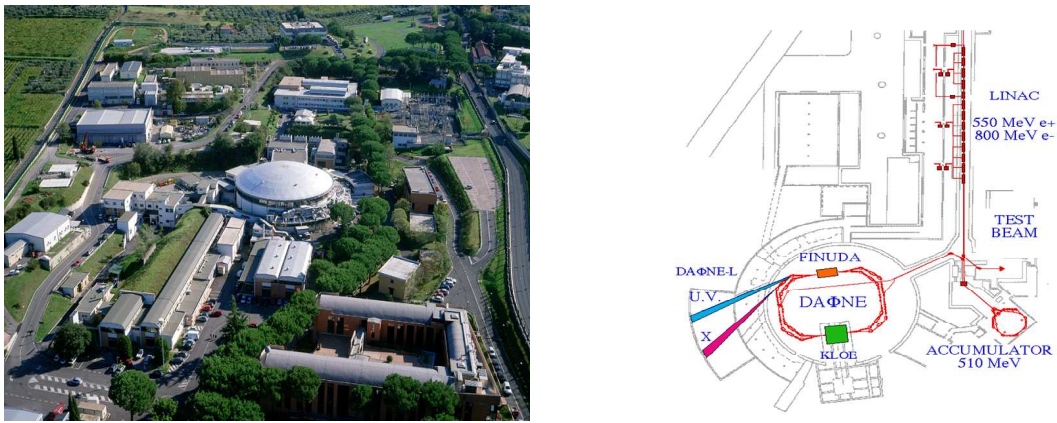

Figure 1. DAΦNE accelerator complex.

Careful beam dynamic studies such as vacuum chamber design with low beam coupling impedance, suppression of different kinds of beam instabilities, investigation of beam-beam interaction, optimization of the beam nonlinear motion have been the key ingredients that have helped to reach this impressive result. Many novel ideas in accelerator physics have been proposed and/or tested experimentally at DAΦNE for the first time. These are suppression of parasitic beam interactions with current-carrying wires [3], beam collisions with negative momentum compaction factor [4], electron cloud mitigation with dedicated electrodes [5], wigglers with "wiggling" poles [6], strong RF focusing proposal [7] and others. Besides, the innovative concept of nonlinear focusing of colliding bunches, called Crab Waist (CW) collision scheme, has been proposed at LNF INFN [8, 9] and successfully tested at DAΦNE in operational conditions providing luminosity for two different experimental detectors, SIDDHARTA [10] and KLOE-2 [11]. Considering high efficiency of the scheme for increasing collision luminosity and its relative simplicity for implementation several new collider projects have been proposed and are under development at present. These are the SuperKEKB B-factory already under commissioning in Japan [12], the SuperC-Tau factory proposed in Novosibirsk and entered in the short list of Russian mega-science projects [13], the new 100-km electron-positron Future Circular Collider (FCC-ee) under design study at CERN [14], the Higgs Factory CEPC in China [15] and some others.

In this paper we give a brief overview of some most relevant advanced accelerator physics studies performed at DAΦNE

## 2. Low impedance vacuum chamber

Since the very beginning of the DAΦNE project beam instabilities were recognized as one of the most harmful potential dangers for collider performance. For this reason the vacuum chamber was designed taking much care of its beam coupling impedance [16], both broad-band [17] and narrow-band.

New designs and novel ideas were adopted for almost all principal vacuum chamber components: RF cavities [18, 19], shielded bellows [20], longitudinal feedback kickers [21], BPMs [22], DC current monitors, injection kickers, transverse feedback kickers and others [23]. For example, the longitudinal feedback kickers based on the DAΦNE kicker design are routinely used in most high intensity lepton colliders and synchrotron radiation sources.

During the years of collider operation the vacuum chamber has been modified several times with the goal of a further impedance reduction. For example, during the collider shutdown for the SIDDHARTA experiment installation the vacuum chamber was substantially revamped [24]: two new low impedance interaction regions were designed and installed, the new fast injection kickers were implemented, the old bellows were substituted by the new ones, having Ω-shape RF sliding contacts. In addition, ion clearing electrodes that were no longer in use, have been removed from the electron ring [25]. At present the measured longitudinal broadband impedance is 0.45 Ω for the electron ring and 0.36 Ω for the positron one [24] to be compared with 13.6Ω of the EPA [26], the former CERN electron-positron accumulator ring with similar circumference and beam energy.

## 3. Collisions with negative momentum compaction factor

Since there are several potential advantages for beam dynamics and luminosity performance of a collider with a negative momentum compaction factor $\alpha_c$ [27] it was suggested to perform a dedicated machine experiment at DAΦNE with $\alpha_c < 0$. The principal results achieved during the experiment were following [4]:

- The DAΦNE lattice is shown to be very flexible and the machine optics model is proven to be reliable in providing stable collider operation with the momentum compaction factors varying in the range from -0.036 to +0.034.
- With the negative $\alpha_c$ bunches in both rings are shortened by wake fields till the microwave instability threshold as predicted by numerical simulations. Figure 2 shows the positron bunch length (left plots) for the positive (green dots) and negative alfa (blue dots) and respective charge distribution at different bunch intensities for $\alpha_c < 0$ (right plot).
- It was possible to store high bunch currents with the large negative chromaticities. No hard limit has been observed in the multibuch operations. About 1 A stable beam currents were stored in both rings.

- At beam currents up to 300 mA per beam a good specific luminosity was obtained in beam-beam collisions.

However, the microwave instability threshold is found to be somewhat lower with $\alpha_c < 0$. Since the experiment was performed before the extraction of the ion clearing electrodes [25] giving dominant contribution into the beam coupling impedance, the high current collisions were prevented by the microwave instability in the electron ring at that time.

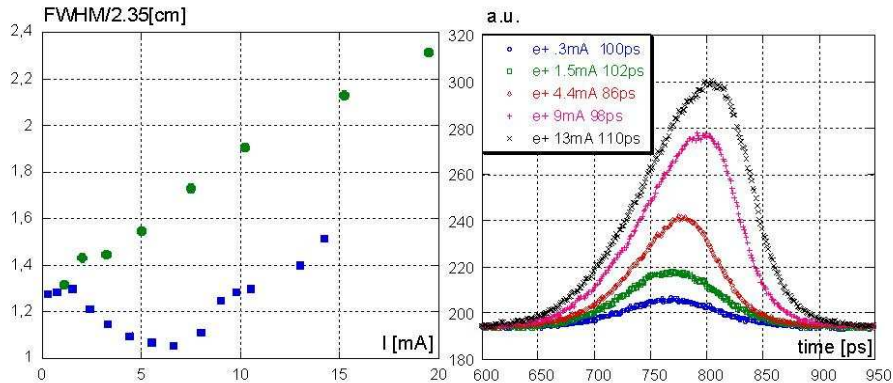

Figure 2 Bunch length (left) and charge distribution (right) for different bunch currents in the DAΦNE positron ring as measured by the streak camera (courtesy A. Stella)

## 4. Parasitic crossing compensation with wires

Long-range beam-beam interactions (parasitic crossings) were one of the sources of luminosity degradation in the DAΦNE original configuration before the collider upgrade for crab waist collisions. Due to drastic lifetime reduction the parasitic crossings put a limit on the maximum storable currents and, as a consequence, on the achievable peak and integrated luminosity. In order to mitigate this problem, numerical and experimental studies of the parasitic crossings compensation by current-carrying wires, the technique originally proposed at CERN [28], have been done [3, 29]. During the operation for the KLOE experiment two such wires were installed at both ends of the interaction region, outside the vacuum chamber. The application of the compensating wires resulted in a two fold improvement of the beam-beam induced losses for identical beam parameters, in agreement with numerical simulations [30].

## 5. e-Cloud clearing electrodes

The maximum electron beam current stored in DAΦNE was 2.5 A, i.e. the record value for the electron beam current ever stored in the modern colliders and synchrotron light sources. However, due to the electron cloud effect [31] we have not been able to exceed 1.4 A in the positron ring. In addition to this current limitation we have been suffering from other harmful e-cloud effects affecting the collider luminosity performance such as anomalous pressure rise, vertical beam size increase, tune spread along the bunch train etc. [32].

In order to cope with the strong e-cloud instabilities powerful bunch-by-bunch feedback systems have been used [33]. In addition, solenoids were installed in order to prevent the e-cloud build-up in the straight sections. However, these measures have not solved completely the problems created by the e-cloud. In particular, the strong horizontal instability still remains the worst trouble for us. Numerical simulations and experimental observations have shown that this kind of instability is triggered by the e-cloud pattern created in the wiggler and dipole magnets [34]. So it has been decided to insert special metallic electrodes [35] to suppress the e-cloud in the dipoles and the wigglers [36]

According to the dedicated numerical simulations application of the DC voltage of 200-500 V at the electrodes should decrease the electron cloud density by about 2 orders of magnitude. Indeed, the electrodes are found to be very useful to reduce the strength of the positron beam horizontal instability; to decrease the betatron tune shift and tune spread inside bunch trains; to suppress the vertical beam size blow-up of the positron beam [5]. As an example, Fig. 3 shows the betatron tune shift and spread reduction with the clearing electrodes switched on.

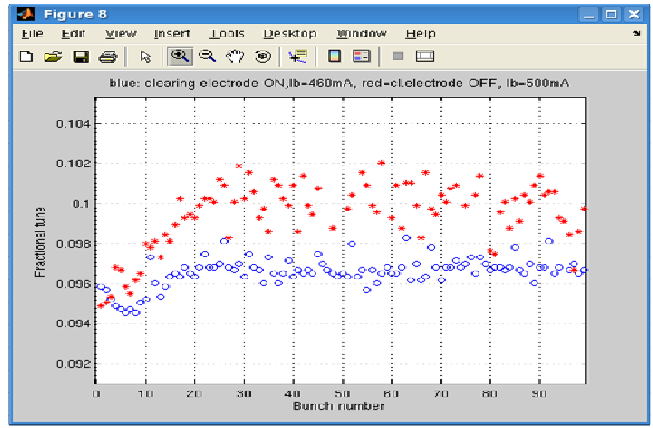

Figure 3 The horizontal betatron tune along the bunch train with the clearing electrodes off (red stars) and on (blue circles) as measured by the transverse feedback system (courtesy A.Drago).

As a result, with the electrodes switched on it is possible to store higher positron beam current, to achieve higher luminosity and to have more stable overall collider performance.

**6. Strong RF focusing**
The longitudinal strong RF focusing is an alternative way to obtain short bunches at the interaction point (IP) [37]. It consists in realizing a magnetic lattice with a large momentum compaction factor together with a strong RF gradient. In this regime the bunch length is no longer constant, but it is modulated along the ring and can be minimized at the IP. In turn, if the main impedance generating elements of the ring are located where the bunch remains long, it is possible to minimize the strength of wake fields. This helps to avoid microwave instabilities and excessive bunch lengthening due to the potential well distortion [7]. The strong RF focusing concept was proposed as one of the possible options for the DAΦNE upgrade [38].

**7. Crab waist collision scheme**
Contrary to the conventional strategies aimed at the luminosity increase, the crab waist collision scheme requires small emittance, large Piwinski angle and larger crossing angle; there is no need to decrease the bunch length and push beam currents beyond the values already achieved in the present factories. This scheme can greatly enhance the luminosity of a collider since it combines several potentially advantageous ideas: collisions with a large Piwinski angle, micro-beta insertions and suppression of beam-beam resonances using the dedicated ("crab waist") sextupoles [39]. For details of beam dynamics in crab waist collision see [9, 40, 41], for example.

The crab waist collision scheme has been successfully tested at the electron-positron collider Φ-factory DAΦNE providing luminosity increase by a factor of 3 for SIDDHARTA experiment [10], in a good agreement with numerical simulations [42]. The simulations have been also very helpful to improve the luminosity and background for the KLOE-2 experiment exploiting the crab waist scheme [43]. Figure 4 demonstrates the effect of the crab sextupoles on the beam-beam blow up and the distribution tails in DAΦNE. The advantages of the crab waist collision scheme have triggered several collider projects exploiting its potential.

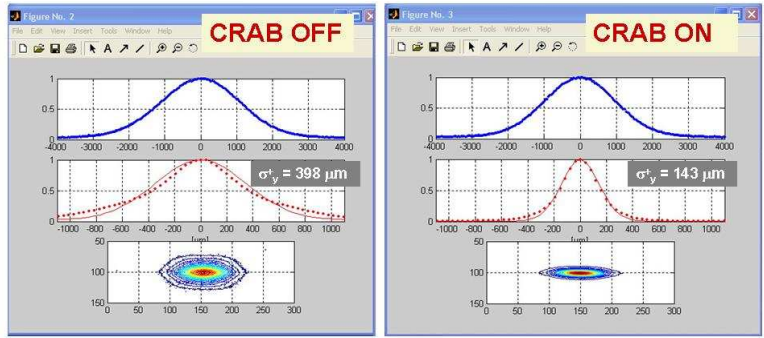

Figure 4. Transverse beam profiles with crab sextupoles off (left) and on (right) as measured by the synchrotron light monitor (courtesy A.Stella)

## 8. Conclusions

After almost 20 years of running DAΦNE is still on track providing high luminosity for physics detectors and performing world-class accelerator experiments. In the future DAΦNE can serve as a test accelerator for new collider projects.

**References**
[1] Vignola G *et al* 1993 DAΦNE, the Frascati Φ-factory *Conf. Proc.* **C930517** pp.1993-1997
[2] Vignola G *et al* 1996 DAΦNE, the first Φ-factory *Conf Proc* **C960610** pp. 22-26
[3] Milardi C *et al* 2006 Wire Compensation of Parasitic Crossings in DAΦNE *Conf Proc* **C060626** pp. 2808-2810 *e-Print:* physics/0607129
[4] Zobov M *et al* 2006 DAΦNE Experience with Negative Momentum Compaction *Conf Proc* **C060626** pp. 989-991 *e-Print:* physics/0607036
[5] Alesini D *et al* 2013 DAΦNE Operation with Electron-Cloud-Clearing Electrodes *Phys. Rev. Lett.* **110** 124801
[6] Bettoni S *et al* 2014 Experimental Validation for the Compensation Method of Nonlinearities in Periodic Magnets, *Phys.Rev.ST Accel.Beams* **17** 102401
[7] Gallo A, Raimondi P and Zobov M 2003 Strong RF Focusing for Luminosity Increase: Short Bunches at the IP, *Conf Proc* **eConf C0309101** THWA003
[8] Raimondi P 2006 Status on SuperB effort *in 2$^{nd}$ SuperB Workshop*, Frascati, Italy
[9] Raimondi P, Shatilov D and Zobov M 2007 Beam-beam issues for colliding schemes with large Piwinski angle and crabbed waist *Preprint* physics/0702033
[10] Zobov M *et al* 2010 Test of crab-wast collisions at DAΦNE Φ-factory *Phys. Rev. Lett.* **104** 174801
[11] Milardi C *et al* 2015 DAΦNE consolidation program and operation with KLOE-2 detector *ICFA Beam Dyn. Newslett.* **67** pp. 9-27
[12] Ohnishi Y *et al* 2013 Accelerator design at SuperKEKB *Prog. Theor. Exp. Phys.* **2013** 03A011
[13] Bondar A *et al* 2013 Project of Super Charm-Tau factory at the Budker Institute of Nuclear Physics *Phys. Atom. Nucl.* **76** pp. 1072-1085.
[14] K.Oide et al 2016 Design of Beam Optics for the Future Circular Collider e+e- Collider Rings *Phys.Rev.Accel.Beams* **17** 102401
[15] Wang D *et al* 2016 CEPC partial double ring scheme and crab-waist parameters *in IAS Program on High Energy Physics* 18-21 January 2016, Hong Kong
[16] Zobov M *et al* 1995 Collective Effects and Impedance Studies for the DAΦNE Φ-Factory, *Proc. Int. Conf. on Collective Effects for B-Factories*, p. 110. *Preprint* LNF-95-041-P.
[17] Bartalucci S *et al* 1993 Broadband Model Impedance for DAΦNE Main Rings *Nucl.Instrum.Meth.* **A337** 231-241

[18] Bartalucci S *et al* 1995 Analysis of Methods for Controlling Multibunch Instabilities in DAΦNE, *Part.Accel.* **48** 213-237
[19] Alesini D *et al* 2004 Third Harmonic Cavity Design and RF Measurements for the Frascati DAΦNE Collider *Phys.Rev.ST Accel.Beams* **7** 092001
[20] Delle Monache G et al 1998 DAΦNE Shielded Bellows *Nucl.Instrum.Meth.* **A403** 185-194
[21] Boni R *et al* 1996 A Waveguide Overloaded Cavity as Longitudinal Kicker for the DAΦNE Bunch-by-Bunch Feedback System *Part.Accel.* **52** 95-113
[22] Marcellini F *et al* 1998 DAΦNE Broad-Band Button Electrodes *Nucl.Instrum.Meth.* **A402** p.27
[23] Zobov M *et al* 1998 Measures to Reduce the Impedance of Parasitic Modes in the DAΦNE Vacuum Chamber Components *Frascati Phys.Ser.* **10** 371-378
[24] Marcellini F *et al* 2008 Coupling Impedance of DAΦNE Upgraded Vacuum Chamber *Conf Proc* **C0806233** TUPP051
[25] Zobov M *et al* 2007 Impact of Ion Clearing Electrodes on Beam Dynamics in DAΦNE *JINST* **2** P08002
[26] Bartalucci S *et al* Collective Effects in the LEP Electron Positron Accumulator (EPA) *Conf. Proc.* **C870316** p. 1234
[27] Ruggiero F and Zobov M 2003 High Luminosity Issues for DAΦNE Upgrade *Conf Proc* **eConf C0309101** SAPL004
[28] Koutchouk J 2000 Principle of a Correction of Long-Range Beam-Beam Effect in LHC Using Electromagnetic Lenses *LHC-Project-Note-223*, CERN.
[29] C.Milardi *et al* 2008 DAΦNE Lifetime Optimization with Compensating Wires and Octupoles *e-Print:* arXiv:0803.1544
[30] A.Valishev et al 2015 Numerical Analysis of Parasitic Crossing Compensation with Wires in DAΦNE *6th Int. Part. Accel.Conf. (IPAC2015)* pp. 589-592 *e-Print:* arXiv:1506.07562
[31] Furman M A 2013 in *Handbook of Accelerator Physics and Engineering* pp 163-166
[32] Drago A, Vaccarezza C and Zobov M 2006 Analysis of Mechanisms Driving the Horizontal Instability in the DAΦNE Positron Ring, *DAΦNE Thechnical Note* G-**67**
[33] Drago A, Teytelman D and Zobov M 2005 Recent Observations on a Horizontal Instability in the DAΦNE Positron Ring in *Proc. of Part. Accel. Conf. (PAC2005),* pp. 1841-1843
[34] Demma T et al 2009 A Simulation Study of the Electron Cloud Instability at DAΦNE in *Proc. of Part. Accel. Conf. (PAC2009),* pp. 4695-4697
[35] Alesini *et al* 2010 Design and Test of the Clearing Electrodes for e-Cloud Mitigation in the e+ DAΦNE Ring *Conf. Proc.* **C100523** TUPEB002
[36] Zobov M *et al* 2012 Operating Experience with Electron Cloud Clearing Electrodes at DAΦNE *Conf. Proc.* **C1206051** pp. 259-265
[37] Gallo A, Raimondi P and Zobov M 2003 Strong RF Focusing for Luminosity Increase *e-Print:* arXiv: physics/0309066.
[38] Gallo A et al 2004 Proposal of a Strong RF Focusing Experiment at DAΦNE in *Proc. of EPAC2004*, pp. 683-685
[39] Raimondi P, Shatilov D and Zobov M 2008 Suppression of beam-beam resonances in crab waist collisions *Conf. Proc.* **C0806233** WEPP045
[40] Zobov M 2011 Next Generation of Electron-Positron Factories *Phys.Part.Nucl.* **42** pp.782-799
[41] Shatilov D, Levichev E, Simonov E and Zobov M 2011 Application of Frequency Map Analysis to Beam-Beam Effects Study in Crab Waist Collision Scheme *Phys. Rev. ST Accel. Beams* **14** 014001
[42] Zobov M 2010 Beam-beam Interaction in Novel, Very High Luminosity Parameter Regimes *Conf. Proc.* **C100523** pp. 3639-3643
[43] Zobov M *et al* 2016 Simulations of Crab Waist Collisions in DAΦNE with KLOE-2 Interaction Region *IEEE Trans.Nucl.Sci.* **63** pp.818-822